\documentclass[epj,draft]{svjour}

\usepackage{epsfig} 
\usepackage{array} 
\usepackage{amssymb}
\usepackage{amsmath} 
\usepackage{amsbsy} 
\usepackage{epsfig}
\usepackage{mathrsfs} 
\usepackage{graphics}
\newcommand{\be}{\begin{equation}}
\newcommand{\ee}{\end{equation}}
\newcommand{\bea}{\begin{eqnarray}}
\newcommand{\eea}{\end{eqnarray}}
\newcommand{\beas}{\begin{eqnarray*}}
\newcommand{\eeas}{\end{eqnarray*}}

\newcommand{\nn}{\nonumber}
\newcommand{\rf}[1]{(\ref{#1})}

\begin{document}

\title{Slip vs viscoelasticity in dewetting thin films}
\titlerunning{Thin-film equations for viscoelastic liquids}

\author{Ralf Blossey\inst{1} \and Andreas M{\"u}nch\inst{2}\and 
Markus Rauscher\inst{3,4} \and Barbara Wagner\inst{5}
%\thanks{\emph{Present address:}}
}                     
%\offprints{}     
%\authorrunning{M. Rauscher, A. M{\"u}nch, P. L. Evans, B. Wagner, and R.
%Blossey}

\institute{
Interdisciplinary Research Institute, c/o IEMN Avenue
Poincar\'e BP 60069, F-59652 Villeneuve d'Ascq, France
\and
Institute of Mathematics, Humboldt University of Berlin,
10099 Berlin, Germany
\and
Max-Planck-Institut f\"ur Metallforschung,
Heisenbergstr. 3, 70569 Stuttgart, Germany 
\and
ITAP, Universit{\"a}t Stuttgart, Pfaffenwaldring 57, 70569 Stuttgart
\and
Weierstrass Institute for Applied Analysis and Stochastics (WIAS),
Mohrenstr. 39, 10117 Berlin, Germany}

\date{Received: date / Revised version: date}

\abstract{Ultrathin polymer films on non-wettable substrates 
display dynamic features which have been attributed to either
viscoelastic or slip effects. Here we show that in the weak 
and strong slip regime effects of viscoelastic relaxation
are either absent or essentially indistinguishable from slip effects. 
Strong-slip modifies the fastest unstable mode in a rupturing
thin film, which questions the standard approach to reconstruct
the effective interface potential from dewetting experiments.}

\PACS{ {83.60.Bc}{Linear viscoelatiscity}    \and
       {47.50.+d}{Non-Newtonian fluid flows} \and 
       {68.15.+e}{Liquid thin films}} 
\maketitle

%%%%%%%%%%%%%%%%%%%%%%%%%%%%%%%%%%%%%%%%%%%%%%%%%%%%%%%%%%%%%%%%%%%%%%%%

%\section{Introduction}
%\label{sec:introduction}
{\it Introduction.} 
In recent years it has been shown that the physics of polymeric thin films 
on non-wettable surfaces can be described, 
to an astonishing level of detail, by 
lubrication models derived from the Navier-Stokes equation for simple 
liquids \cite{seemann01,becker03}. However, ultrathin dewetting films exhibit 
unusual features in their rupture dynamics which show up in the morphology 
and velocities of dewetting holes \cite{reiter01}. It has been 
suggested that viscoelasticity plays an important role in these films, 
in particular when the polymer length scales become comparable to the film 
thickness. There is now a large number of modelling attempts
to explain these features 
\cite{herm02,saulnier02,shenoy02,herm03,vilmin05-2}. 
Most of them assume a generalized Maxwell- or Jeffreys-type 
dynamics for the stress-strain relation in these films, sometimes in 
combination with additional flow functions; all these assumptions are, while 
not entirely artificial, hard to soldily justify at present. In the absence of
better knowledge, the Jeffreys-model therefore remains a useful starting point 
for modeling, with the idea to confront the predictions with experiment.

As has been shown very recently, thin-film lubrication models can be
classified into different slip-classes, and separate models have to
be derived for each class. There are models valid specifically in the
limit of strong slip \cite{kargupta04,muench05} but also in
intermediate slip regimes \cite{muench05}. The distinction of different 
slip classes is essential for the description of dewetting
experiments of PS-films on substrates with different slip properties
\cite{fetzer05}. This last result has shown that slip effects can indeed
explain the anomalies in the shape of dewetting films. 

As we demonstrate here, the distinction of different slip classes 
remains true for viscoelastic thin films of Jeffreys type. We show that 
it is easy to generalize the recently proposed lubrication model for 
Newtonian liquids in the strong-slip regime \cite{kargupta04,muench05}
to a Jeffreys model. We here apply this model, as well as the recently
developed model for the weak slip case \cite{rauscher05}, to determine
the onset conditions of rupture in unstable thin films. 
\\

{\it Model assumptions and lubrication equations.} 
We begin with the bulk dynamic equations for the viscoelastic liquid. 
It is assumed as incompressible, hence the velocity field ${\bf u} = 
(u_x, u_y, u_z) $ fulfills the mass conservation equation
\begin{equation}  \label{one}
{\bf \nabla} \cdot {\bf u} = 0\, .
\end{equation}
The equation of momentum conservation is given by
\begin{equation} \label{two}
\varrho \frac{d{\bf u}}{dt} = - {\bf \nabla}p_R + {\bf \nabla} \cdot {\bf \tau}
\end{equation}
where $p_R = p + V' $ is the augmented pressure, with $p$ as capillary pressure
and $V'$ as the van der Waals-type dispersion forces. The traceless part of 
the stress tensor is described by a symmetric matrix $\tau$. Further, in 
eq.(\ref{two}), $d/dt$ is the total derivative, and $\nabla$ abbreviates 
the partial derivative vector with entries $\partial_i $, $i = x,y,z $.

To complete the model we have to choose a constitutive relation for the
stress tensor $\tau$. As argued in the introduction, we opt for the linear 
Jeffreys model defined by
\begin{equation}
(1 + \lambda_1 \partial_t ){\bf \tau} = \eta(1
+ \lambda_2 \partial_t) {\bf \dot{\gamma}}
\end{equation}
in which ${\bf \dot{\gamma}} $ is the strain rate,
$\dot{\gamma}_{ij} = \partial_i u_j + \partial_j u_i $.
The rates $\lambda_1$, $\lambda_2$ govern the relaxation of the stress
and strain rate, respectively.

In order to derive the equations for a thin film of height $z = h(x,y,t) $
we have, for the incompressible case, the kinematic condition
\begin{equation}
\partial_t h = - \nabla_{\parallel}\cdot\int_0^h dz\,\, {\bf u}_{\parallel} 
\end{equation} 
where $\nabla_{\parallel} = (\partial_x,\partial_y) $, and $ {\bf u}_{\parallel}
= (u_x,u_y) $.
The boundary conditions at the free surface correspond to the vanishing
of the stress tensor components tangential to the film surface (i.e.,
we neglect the vapor phase), 
while the normal component of the stress tensor obeys
\begin{equation}  \label{surf}
({\bf \tau} - p_R{\bf 1})\cdot {\bf n} = 2 \sigma \kappa {\bf n}\, 
\end{equation}
where $\sigma $ is the surface tension of the film, ${\bf 1} $ a $ 3 \times
3 $ unit matrix, and $\kappa$ the local mean curvature with sign convention
that $\kappa < 0 $ for a spherical drop. Finally, in eq.(\ref{surf}), 
the normal vector to the film is given by 
${\bf n} = (-\nabla_{\parallel}h,1)/\sqrt{g}$, 
with $ g = 1 + (\nabla_{\parallel} h)^2 $.
The model is completed by the boundary conditions at the surface which 
are of Navier type, i.e.,
\begin{equation}
u_z = 0\,,\,\,\,\,\,\,\, u_i = \frac{b}{\eta}\tau_{iz}
\end{equation}
where $b$ is the slip length.

We now sketch the derivation of the lubrication model for strong and
weak slip that can be derived from this bulk dynamics; the details
can be found in refs. \cite{rauscher05,muench05}, and in the Appendix
to this paper.

First, we introduce a measure of the relative scale of the thin film
height $H$, $ z = H z^* $, to its lateral extension, $L$, $ (x,y) =
L (x^*,y^*) $, and we define $\epsilon \equiv H/L \ll 1$. Time is scaled 
by $ T = U/L$, where $U$ is the velocity scale. The stress tensor scales as
\begin{equation}
\tau_{ij} = \frac{\eta}{T}\tau_{ij}^*
\end{equation}
for $(i,j) = (x,y) $ and, additionally, $i=j$. The remaining components
scale as
\begin{equation}
\tau_{ij} = \frac{\eta}{\varepsilon T} \tau_{ij}\, .
\end{equation}
The distinction between weak and strong slip lengths arises from the
choice of balancing conditions between the forces acting on the film.
In the weak slip limit, one has with the pressure scale $P$ \cite{rauscher05}
\begin{equation}
\frac{PH}{\eta U} \sim \varepsilon^{-1}\,,
\end{equation}
while in the strong slip limit we need \cite{muench05}
\begin{equation}
\frac{PH}{\eta U} \sim \varepsilon\, .
\end{equation}
The Reynolds number $Re = \varrho U L/\eta $ now scales as either
\begin{equation}
Re = \epsilon^3 Re^*
\end{equation}
in the weak slip case, or as
\begin{equation}
Re = \epsilon Re^*
\end{equation}
in the strong slip case, where $ Re^*$ is the reduced Reynolds number of 
order one. Finally, the slip length $b$ scales as $ b = O(1)$ 
in the weak slip regime and $b=\beta_s\epsilon^{-2}$ in the strong slip case.

We first state the result for the strong slip case, details are given in the
Appendix. Being interested here only in the conditions of thin film rupture, 
we restrict the discussion to the (laterally) one-dimensional case; 
the extension to the full two-dimensional case is straightforward.

In the strong slip lubrication limit one ends up with the following
system of equations (we put $\sigma=1$),
\begin{eqnarray} \label{sslip}
h Re^* (\partial_t u + u \partial_x u) & = &
h \partial_x [\partial^2_{x}h - V'(h)] + \partial_x (4hq) - \frac{u}{\beta_s}\,,
\nonumber \\ 
\nonumber \\
(1 + \lambda_1\partial_t)q & = & (1 + \lambda_2\partial_t)\partial_x u\,,
\\
\nonumber \\
\partial_t h + \partial_x (hu) & = & 0\,. \nonumber
\end{eqnarray}
where $q$ is related to the stress tensor, see Appendix.
Note that the system (\ref{sslip}) readily reduces to the Newtonian case if
$\lambda_1 = \lambda_2 = 0$; the added complexity of the viscoelasticity
is thus relatively minor in this limit.

By contrast, in the weak-slip limit, one is able to derive the equation
\cite{rauscher05}
\begin{eqnarray} \label{weak}
& & (1 + \lambda_2\partial_t)\partial_t h +  (\lambda_2 - \lambda_1)\partial_x
\left(\frac{h^2}{2}Q - hR\right) \partial_t h= \\ \nonumber
&& -\partial_x \left[\left((1 + \lambda_1\partial_t)\frac{h^3}{3}
+ (1 + \lambda_2 \partial_t)bh^2 \right)\partial_x (\partial_x^2 h-V'(h))\right]\,,
\end{eqnarray}
where
\begin{equation}
(1 + \lambda_2\partial_t)Q = -\partial_x (\partial_x^2 h-V'(h))
\end{equation}
and
\begin{equation}
(1 + \lambda_2\partial_t)R = -h \partial_x (\partial_x^2 h-V'(h))\, .
\end{equation}
Note that for $\lambda_2 \rightarrow 0 $, eq.(\ref{weak}) collapses
to a single equation; this limit corresponds to the simplest Maxwell
model. In the case $\lambda_1 = \lambda_2 $ one recovers the thin-film
equation of the Newtonian liquid with an extra multiplicative factor
$ (1 + \lambda_1\partial_t) $ on both sides.
\\

{\it Linear stability analysis.}
We now turn to the linear stability analysis of a thin film which 
experiences a dispersion forces which destabilizes it (i.e., $ V''(h_0) < 0 $).
The two different cases yield:
\\

A) {\it Weak slip.} The linear stability analysis is easily determined by 
assuming 
\begin{equation}
h = h_0 + \delta h_1\,,\,\,\,\,\, Q = \delta Q_1\,,\,\, R = \delta R_1,
\end{equation}
where $0<\delta\ll 1$ 
with, in addition
\begin{equation} \label{mode}
 (h_1(x,t), Q_1(x,t),R_1(x,t))\equiv (\hat h_1,\hat Q_1,\hat R_1) e^{ikx + \omega t}\, .
\end{equation}
The resulting dispersion relation $\omega(k)$ can be expressed as
\begin{equation} \label{weakdisp}
(1 + \lambda_2 \omega) \omega = \omega_N(1 + \Lambda\omega)
\end{equation}
where 
\begin{equation}
\omega_N(k) = - \left(\frac{h_0^3}{3} + bh_0^2\right)g(k)
\end{equation}
with
\begin{equation}
g(k) = k^4 + k^2 V''(h_0)
\end{equation}
is the dispersion relation of the Newtonian liquid, and
\begin{equation}
\Lambda \equiv \lambda_2 + 
\frac{(\lambda_1 - \lambda_2)h_0^3}{h_0^3 + 3bh_0^2}
\end{equation}
From eq.(\ref{weakdisp}) it is easy to see that the structure of the
dispersion relation of the Jeffreys film is identical to that of the
Newtonian film. The range of unstable modes is the interval between
the two zeroes of eq.(\ref{weakdisp}) which is given by the two zeroes
of $g(k)$. Further, also the fastest unstable mode is unaffected by 
viscoelastic relaxation.
\\

B) {\it Strong slip.} In complete analogy to case a) one puts
\begin{equation}
h = h_0 + \delta h_1\,,\,\,\,\,\, q = \delta q_1\,,\,\, u = \delta u_1
\end{equation}
and, with eq.(\ref{mode}), one finds the dispersion relation
\bea
&&(1 + \lambda_1 \omega) (h_0 Re^* \omega + \beta_s^{\, -1}) \omega
+ 4 h_0 k^2 \omega(1 + \lambda_2 \omega)\nn\\
&& + h_0^2 g(k) (1 + \lambda_1 \omega)= 0\, .
\eea
Again it is immediately evident that the range of unstable modes is
unaffected by viscoelastic relaxation. 
The most unstable wavenumber $k_m$ satisfies for the case Re$^*=0$
which applies to the systems studied in \cite{fetzer05}
\bea\label{km}
4\beta_s h_0^3 k_m^4
+h_0^2\left(2k_m^2+V''(h_0)\right)\frac{1+\lambda_1\beta_s h_0^2k_m^4}{1+\lambda_2\beta_s h_0^2k_m^4}=0.
\eea
This result shows that the most unstable mode is strongly affected
by slip, as was already observed for the case of a Newtonian liquid 
\cite{kargupta04}. In addition we find that $k_m$ also depends 
on the relaxation parameters $\lambda_1$ and $\lambda_2$. 
In the limit $\lambda_1$, $\lambda_2\gg 1$ eq. \rf{km} 
simplifies to 
\be
k_m^2=-\frac{\rho}{4}\pm\sqrt{\frac{\rho^2}{16}-\frac{V''(h_0)\rho}{4}},\quad
\rho=\frac{\lambda_1}{\beta_s h_0 \lambda_2}\, .
\ee
This result also holds for Re$^*\neq 0$, but the condition 
corresponding to eq.\rf{km} is much more involved. 
\\ 

{\it Conclusion.} Based on the derivation of lubrication models for 
thin-film dynamics of Jeffreys type we conclude that both in the
weak and strong slip limits, linear viscoelastistic effects are
essentially absent for film rupture. By contrast, strong slip affects
the the most preferred wavenumber, which now also depends on the 
relaxation parameters. In particular, from eq.(\ref{km}) it appears 
that the standard approach for the reconstruction of the
interface potential, which is based on the wavelength of the fastest 
growing mode is questionable for films subject to strong slip.
\\

{\it Appendix A: Strong-slip lubrication limit for the Jeffreys model.}
\\
 
The derivation of the strong-slip lubrication model for the linear
Jeffreys case follows closely both the calculation in the weak slip
regime, and the strong-slip Newtonian case. As in \cite{muench05},
the starting point is the ansatz
\begin{eqnarray}
(u,w,h,p_R,\tau_{ij}) & = & (u_0,w_0,h_0,p_{R0},\tau_{ij0}) \\
& + & \epsilon^2 (u_1,w_1,h_1,p_{R1},\tau_{ij1}) \nonumber
\end{eqnarray}
where $u$ and $w $ are the velocity field components in $x$ and $z$-directions,
neglecting the transverse $y$-direction.
To leading order we find the equations
\begin{equation}
\tau_{xz0} = 0 
\end{equation}
\begin{equation}
(1 + \lambda_2\partial_t)\partial_z u_0 = 0
\end{equation}
with the solution
\begin{equation}
\partial_z u_0 = c(x,z)\exp(-t/\lambda_2)\, .
\end{equation}
We select the solution $ c \equiv 0 $ since any other solution would 
correspond to a strong prestressing of the film at times $t \rightarrow
-\infty $. Therefore, $ u_0 = f(x,t) $, and from the mass conservation 
we have $ \partial_x f = - \partial_z w_0 $, hence $ w_0 = -z \partial_x f $.
It thus follows
\begin{equation}
(1 + \lambda_1 \partial_t)\tau_{zz0} = 
-(1 + \lambda_2 \partial_t)\partial_x f\, .
\end{equation} 
which reads in integrated form as
\begin{equation}
\tau_{zz0} = - \frac{2}{\lambda_1}\int_{-\infty}^t 
dt' e^{(t - t')/\lambda_1}(1 + \lambda_2 \partial_t)\partial_x f 
= - \tau_{xx0}\, .
\end{equation}
To solve for $f(x,t)$, we need to make use of the next order, which
gives
\begin{equation}
Re^*(\partial_t + f\partial_x) f = \partial_x \tau_{xx0} + \partial_z 
\tau_{xz1} = - \partial_x p_{R0}
\end{equation} 
where $ p_{R0} = - \partial_{xx}h_0 - \tau_{zz0}\,$.
This can be written as
\begin{eqnarray} \label{req}
Re^*(\partial_t + f\partial_x) f & = & \partial_z \tau_{xz1} + 
\partial_{xxx}h_0 \\ 
& + & \frac{4}{\lambda_1}\partial_x  \int_{-\infty}^t
dt' e^{(t - t')/\lambda_1}(1 + \lambda_2 \partial_t)\partial_x f \nonumber
\end{eqnarray}
From the boundary condition at the free surface we find to second order
\begin{equation}
((\tau_{xx0} - \tau_{zz0}) + \tau_{xz0}(\partial_x h_0))(\partial_x h_0)
= \tau_{xz1}
\end{equation}
and hence 
\begin{equation}
\tau_{xz1} = - 2 (\partial_x h_0) \tau_{zz0}\, .
\end{equation}
It remains to determine the second order result from boundary condition 
at the substrate. We have $ \tau_{xz1} = f/\beta $ and can
now integrate eq.(\ref{req}) with respect to $ z$ across the film from
$ 0 $ to $h_0 $ and obtain the system of eqs. given in the text
(with $ f \equiv u $), and where 
\begin{equation}
q = - \frac{\tau_{zz0}}{2}\, .
\end{equation}

\end{document}